\documentclass[conference]{IEEEtran}
\IEEEoverridecommandlockouts
\usepackage{booktabs}
\usepackage{lipsum}
\usepackage{xcolor,soul,framed} 

\usepackage{diagbox}
\usepackage{array}
\usepackage{kantlipsum}
\usepackage{caption}
\usepackage{subcaption}

\usepackage{xcolor,soul,framed} 

\colorlet{shadecolor}{yellow}
\usepackage[pdftex]{graphicx}
\graphicspath{{../pdf/}{../jpeg/}}
\DeclareGraphicsExtensions{.pdf,.jpeg,.png}
\usepackage{multirow}
\usepackage[cmex10]{amsmath}
\usepackage{pgfplots}
\pgfplotsset{compat=1.17}

\usepackage{array}
\usepackage{mdwmath}
\usepackage{mdwtab}
\usepackage{eqparbox}
\usepackage{url}
\usepackage{capt-of}  
\usepackage{cuted}    
\begin{document}


\title{Enhancing IoT Intrusion Detection Systems through Adversarial Training\\
}

\author{\IEEEauthorblockN{1\textsuperscript{st} Karma Gurung }
\IEEEauthorblockA{\textit{Computer Science and Engineering} \\
\textit{Wright State University}\\
Ohio, USA \\
gurung.43@wright.edu}
\and
\IEEEauthorblockN{2\textsuperscript{nd} Ashutosh Ghimire }
\IEEEauthorblockA{\textit{Computer Science and Engineering} \\
\textit{Wright State University}\\
Ohio, USA \\
ashutosh.ghimire@wright.edu}
\and
\IEEEauthorblockN{3\textsuperscript{rd}  Fathi Amsaad}
\IEEEauthorblockA{\textit{Computer Science and Engineering} \\
\textit{Wright State University}\\
Ohio, USA \\
fathi.amsaad@wright.edu}
}

\maketitle

\begin{abstract}
The  augmentation of Internet of Things (IoT) devices transformed both automation and connectivity but revealed major security vulnerabilities in networks. We address these challenges by designing a robust intrusion detection system (IDS) to detect complex attacks by learning patterns from the NF-ToN-IoT-v2 dataset. Intrusion detection has a realistic testbed through the dataset's rich and high-dimensional features. We combine distributed preprocessing to manage the dataset size with Fast Gradient Sign Method (FGSM) adversarial attacks to mimic actual attack scenarios and XGBoost model adversarial training for improved system robustness. Our system achieves 95.3\% accuracy on clean data and 94.5\% accuracy on adversarial data to show its effectiveness against complex threats. Adversarial training demonstrates its potential to strengthen IDS against evolving cyber threats and sets the foundation for future studies. Real-time IoT environments represent a future deployment opportunity for these systems while extensions to detect emerging threats and zero-day vulnerabilities would enhance their utility.
\end{abstract}

\section{Introduction}

The application of the Internet of Things (IoT) has revolutionized industries such as healthcare, agriculture, smart homes, transportation,  and smart cities by enabling interconnected devices to communicate and collaborate seamlessly. This advancement has improved operational efficiency and facilitated data-driven decision-making. However, the rapid increase in IoT devices has introduced significant security challenges. The heterogeneous nature of IoT devices, coupled with their limited computational resources and dynamic deployment environments, makes them highly susceptible to cyber threats. Attack vectors such as Distributed Denial-of-Service (DDoS) attacks, ransomware, injection attacks, and cross-site scripting (XSS) can compromise data integrity, disrupt critical services, and result in financial and reputational damages~\cite{sarhan2020, sarhan2021feature}.

To address these challenges, robust Intrusion Detection Systems (IDS) have become an indispensable component of IoT network security. However, traditional IDS solutions often fail to meet the unique demands of IoT environments, which require systems to detect diverse and evolving attack patterns while maintaining computational efficiency. Machine learning-based IDS have gained prominence for their ability to learn complex patterns from network traffic data and adapt to new attack types~\cite{zhang2024egracl, aljamal2024robust}. Despite their advantages, these systems remain vulnerable to adversarial attacks, where malicious perturbations in the input data can lead to misclassification~\cite{goodfellow2015, raskovalov2022}.

This study focuses on developing a robust IDS by leveraging the NF-ToN-IoT-v2 dataset, a comprehensive and high-dimensional dataset specifically designed for IoT network intrusion detection~\cite{sarhan2020}. The proposed IDS employs the XGBoost machine learning algorithm, known for its scalability, computational efficiency, and superior performance in handling structured data~\cite{chen2016xgboost}. To enhance resilience against adversarial attacks, the study integrates adversarial training into the model, ensuring improved robustness in detecting sophisticated intrusions.

Key contributions of this work include the implementation of scalable preprocessing techniques using distributed computing to handle the large and complex NF-ToN-IoT-v2 dataset efficiently~\cite{dask2015parallel}. The resilience of the IDS is evaluated using adversarial examples generated through the Fast Gradient Sign Method (FGSM), which simulate real-world attack scenarios and expose potential vulnerabilities~\cite{goodfellow2015}. To mitigate these vulnerabilities, adversarial training is incorporated into the XGBoost model, enabling it to detect intrusions effectively even in the presence of adversarial perturbations. Extensive experiments demonstrate the model’s effectiveness, achieving a classification accuracy of 95.3\% on clean test data and 94.5\% on adversarial examples. These results underscore the robustness and reliability of the proposed IDS in detecting a wide range of attack types.

Since Section II examines relevant work in IoT intrusion detection and adversarial training, the remainder of the study is divided into various chapters. The NF-ToN-IoT-v2 dataset and preparation methods are described in depth in Section III. The methodology, including the use of training and adversarial attacks, is explained in Section IV. The results of the experiment and an assessment of the suggested methodology are presented in Section V. The work is finally concluded in Section VI, which also offers ideas for future research possibilities.

\section{Related Work}

\subsection{IoT Intrusion Detection}

Intrusion detection systems (IDS) have emerged as a critical component of IoT network security due to the increasing frequency and complexity of cyberattacks targeting IoT devices. The NF-ToN-IoT dataset has established itself as a benchmark for evaluating machine learning-based IDS by providing a rich and diverse representation of IoT network traffic~\cite{sarhan2020}.

Sarhan et al.~\cite{sarhan2020} introduced the NF-ToN-IoT dataset, which improved upon its predecessor, the ToN-IoT dataset, by presenting features in NetFlow format. This format facilitates efficient feature extraction and evaluation for network intrusion detection tasks, particularly for complex attacks such as DDoS and ransomware. The dataset has since become a cornerstone for benchmarking machine learning models in the IoT domain.

Further advancing the dataset’s utility, Sarhan et al.~\cite{sarhan2021feature} conducted a detailed feature analysis to identify critical attributes that contribute significantly to intrusion detection. Their work optimized feature selection, which not only improved model accuracy but also reduced computational overhead, a key consideration for IoT environments. Raskovalov et al.~\cite{raskovalov2022} addressed inconsistencies in attack labels and feature representations within the dataset, proposing a standardized feature set that enhances its applicability for advanced models such as Graph Neural Networks (GNNs). These rectifications have significantly expanded the dataset's usability for modern machine learning approaches.

Zhang et al.~\cite{zhang2024egracl} applied GNNs to the NF-ToN-IoT dataset, demonstrating the effectiveness of graph-based methods for capturing complex relationships between network events. Their results highlighted the superior performance of GNNs in multi-class classification tasks, paving the way for incorporating graph-based techniques into IDS. Despite these advancements, limited attention has been given to the robustness of IDS against adversarial attacks.

Despite these advancements, limited attention has been given to the robustness of IDS against adversarial attacks.
Additionally, while most studies focus on supervised learning-based intrusion detection, there is growing interest in unsupervised anomaly detection methods that can operate effectively with limited labeled data. For instance, Ghajari et al. ~\cite{ghajari2024hybrid} proposed a hybrid unsupervised anomaly detection framework that integrates distance and local density measures to enhance early detection capabilities in data-scarce environments. Although originally applied to pandemic case identification, such hybrid approaches offer valuable insights for designing scalable and resilient IDS for IoT networks, where early detection of novel threats and zero-day attacks remains a critical challenge.

\subsection{Adversarial Robustness}

Adversarial attacks pose a significant challenge to machine learning-based IDS by introducing subtle perturbations to input data, which can lead to misclassification~\cite{goodfellow2015}. Among these, the Fast Gradient Sign Method (FGSM) has been widely studied for its ability to generate adversarial examples efficiently and expose vulnerabilities in model decision boundaries. 

AlJamal et al.~\cite{aljamal2024robust} investigated the impact of adversarial attacks on IoT networks, focusing on detecting XSS attacks. They employed adversarial training and achieved a detection accuracy of 99.89\% on the NF-ToN-IoT-v2 dataset, showcasing the potential of this defense mechanism in improving IDS robustness. Other studies have demonstrated that adversarially trained models are better equipped to handle perturbed inputs, although this often comes at the cost of slightly reduced accuracy on clean data~\cite{madry2018}.

Raskovalov et al.~\cite{raskovalov2022} explored the implications of adversarial robustness for GNN-based IDS, proposing a standardized feature set for mitigating adversarial vulnerabilities. Their findings underscore the importance of adapting datasets and models to better withstand adversarial scenarios, particularly in the context of IoT security.

Building upon these foundational works, our research integrates FGSM-based adversarial attacks with XGBoost to evaluate and enhance the robustness of IDS against a broader spectrum of IoT-specific threats. Unlike prior studies that primarily focus on binary classification tasks, our approach addresses multi-class classification challenges.

\subsection{Summary of Related Work}

The existing literature underscores the NF-ToN-IoT dataset's pivotal role in advancing IoT security research. While substantial progress has been made in optimizing feature selection~\cite{sarhan2021feature}, leveraging modern architectures like GNNs~\cite{zhang2024egracl}, and addressing adversarial vulnerabilities~\cite{aljamal2024robust}, challenges persist in ensuring robust model performance under adversarial conditions. Our work bridges this gap by integrating adversarial training into traditional IDS methodologies, offering a comprehensive solution for enhancing IDS resilience in IoT networks.

\section{Dataset and Preprocessing}

\subsection{NF-ToN-IoT-v2 Dataset}

The NF-ToN-IoT-v2 dataset~\cite{alsaedi2020ton_iot} has been specifically designed to support research into intrusion detection in IoT networks and is directly prepared for that purpose. It is based on the NF-ToN-IoT dataset~\cite{sarhan2020}, with more precise labeling and additional features, which makes it appropriate for multi-class classification tasks. It has more than 13 million records and includes a large range of network traffic activities with both normal and abnormal traffic, including types of attack such as benign, backdoor, DDoS, dos, injection, MITM, password, ransomware, scanning, and XSS.

\begin{figure}[t]
    \centering
    \includegraphics[width=0.45\textwidth]{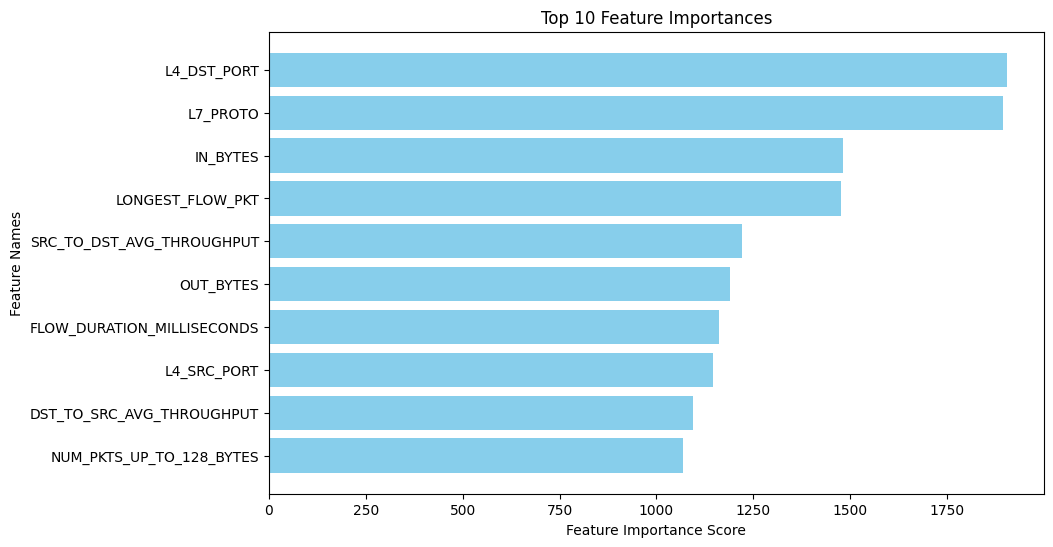} %
    \caption{Top 10 important features}
    \label{feature_importance}
\end{figure}

Key features ~\ref{feature_importance} of the dataset includes network attributes (e.g., source and destination IP addresses, ports, and protocols), traffic statistics (e.g., packet counts, payload sizes, and flow durations), and anomaly labels that classify network traffic into specific types of malicious and benign behaviors. These features provide a realistic representation of IoT network traffic, enabling robust model evaluation~\cite{sarhan2021feature, raskovalov2022}.

The combination of high-dimensional numerical and categorical features makes the NF-ToN-IoT-v2 dataset a robust benchmark for evaluating intrusion detection systems (IDS). Its design facilitates scalability and adaptability, addressing the evolving threat landscape in IoT environments~\cite{zhang2024egracl}~\ref{feature_importance}.

\subsection{Distributed Preprocessing}

The significant scale and complexity of the NF-ToN-IoT-v2 dataset required a scalable and efficient preprocessing pipeline. To address this, we utilized Dask, a distributed computing framework, which is well-suited for handling large-scale datasets in parallel environments~\cite{rocklin2015dask}. Dask allowed us to perform preprocessing tasks efficiently, reducing computational overhead and ensuring the integrity of the data.

Key preprocessing steps included:

1. \textbf{Handling Missing Values:} Numerical features with missing values were imputed using their mean, while categorical features were assigned placeholders. This approach ensured data completeness without introducing bias~\cite{kuhn2013applied}.

2. \textbf{Data Normalization:} Numerical features were normalized using min-max scaling to standardize the dataset. During training, normalization stopped characteristics with higher magnitudes from unduly affecting the model.~\cite{jain2005data}.

3. \textbf{Label Encoding:} Attack labels, initially represented as strings (e.g., "ransomware," "XSS"), were converted into numerical classes using label encoding. This transformation ensured compatibility with machine learning algorithms, particularly for multi-class classification tasks~\cite{bishop2006pattern}.

By implementing these preprocessing steps in a distributed manner, we achieved significant efficiency gains, allowing us to prepare the dataset for subsequent training and evaluation without sacrificing quality.

\subsection{Dataset Splitting}

To enable effective training and evaluation, the preprocessed dataset was divided into training and testing subsets using a 70-30 split. This split resulted in approximately 9.1 million records for training and 3.9 million for testing, ensuring that the model had access to a diverse set of patterns for learning while retaining a robust test set for evaluation. 

The large training set enabled the model to generalize across different attack patterns, while the substantial test set provided a comprehensive platform for assessing the model's performance under realistic conditions. This approach aligns with best practices in machine learning for handling large-scale datasets~\cite{friedman2001elements}.

The use of Dask for distributed preprocessing further enhanced scalability, reducing computation time and enabling real-time adjustments to the data pipeline. This preprocessing framework ensures that the dataset retains its diversity and integrity, making it suitable for advanced machine learning and adversarial training experiments.

\section{Methodology}

This section outlines the methods employed to develop a robust Intrusion Detection System (IDS) using the NF-ToN-IoT-v2 dataset. The methodology encompasses the implementation of the XGBoost model, adversarial attacks using the Fast Gradient Sign Method (FGSM), and adversarial training to enhance the model’s resilience.

\subsection{XGBoost Model}

XGBoost, an advanced gradient-boosting algorithm, was chosen for its superior performance on large-scale, high-dimensional datasets~\cite{chen2016xgboost}. The model's ability to handle missing values, regularization techniques to prevent overfitting, and efficient parallel processing capabilities make it particularly suited for the NF-ToN-IoT-v2 dataset~\cite{chen2020systematic}. 

The model was configured with the following hyperparameters:
\begin{itemize}
    \item \textbf{Objective:} Multi-class classification to predict multiple types of network attacks.
    \item \textbf{Evaluation Metric:} Logarithmic loss (log-loss) for multi-class classification, which measures the accuracy of probabilistic predictions.
    \item \textbf{Tree Depth:} A maximum depth of 5 to balance model complexity and prevent overfitting.
    \item \textbf{Learning Rate:} Set to 0.1, enabling the model to converge steadily while maintaining accuracy.
\end{itemize}

The computational efficiency of XGBoost was further enhanced by employing GPU acceleration through the \texttt{tree\_method="gpu\_hist"} option. GPU-based parallel processing significantly reduced the training time, allowing for faster experimentation and iterative model optimization~\cite{rana2020accelerating}.

\subsection{FGSM Adversarial Attack}

The Fast Gradient Sign Method (FGSM) is a widely used technique to evaluate model robustness by generating adversarial examples~\cite{goodfellow2014explaining}. These examples are crafted by introducing small perturbations to input data, designed to maximize the model's prediction loss. The adversarial examples were generated using the following formula:

\begin{equation}
X_{\text{adv}} = X + \epsilon \cdot \text{sign}\left(\nabla_X J(\theta, X, y)\right)
\end{equation}

where:
\begin{itemize}
    \item \( X \) represents the original input data.
    \item \( \epsilon \) is the perturbation magnitude, controlling the severity of the attack.
    \item \( \nabla_X J(\theta, X, y) \) is the gradient of the loss function \( J \) with respect to the input \( X \), computed at model parameters \( \theta \).
    \item \( \text{sign}(\cdot) \) denotes the element-wise sign function.
\end{itemize}

FGSM was applied to the clean test data from the NF-ToN-IoT-v2 dataset to simulate real-world adversarial attack scenarios. These perturbations exposed vulnerabilities in the model's decision boundaries, providing insights into its susceptibility to adversarial threats~\cite{kurakin2018adversarial}.

\begin{figure}[!t]
    \centering
    \includegraphics[width=\linewidth]{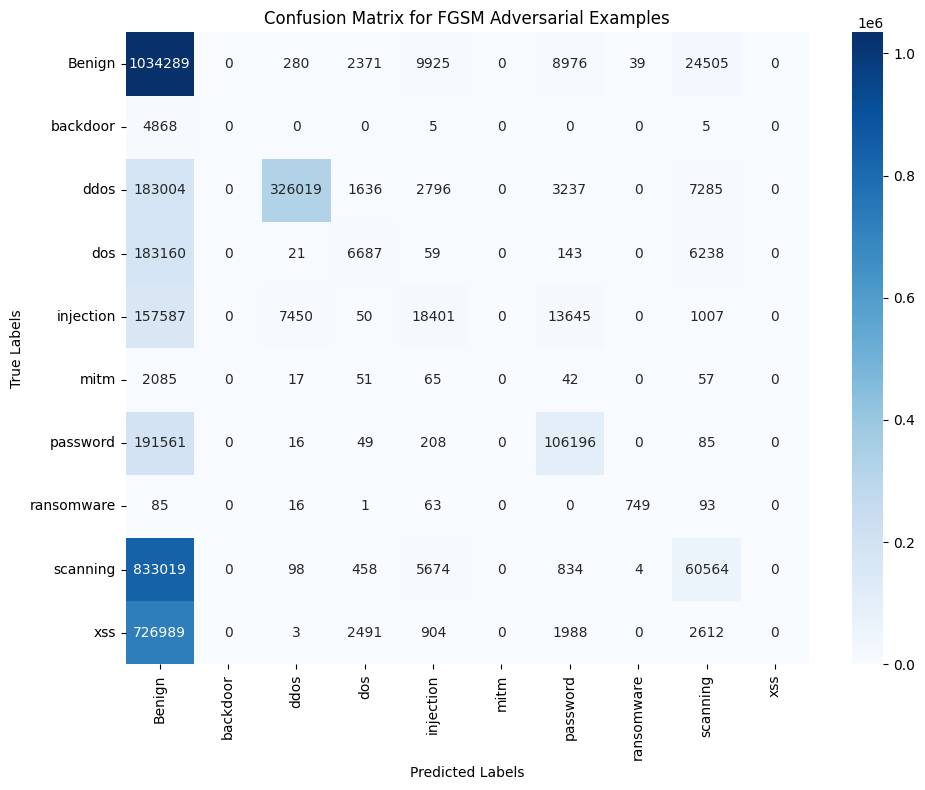} 
    \caption{Confusion Matrix for Adversarially Trained.}
    \label{CM with adversarially trained}
\end{figure}

\begin{figure}[!t]
    \centering
    \includegraphics[width=\linewidth]{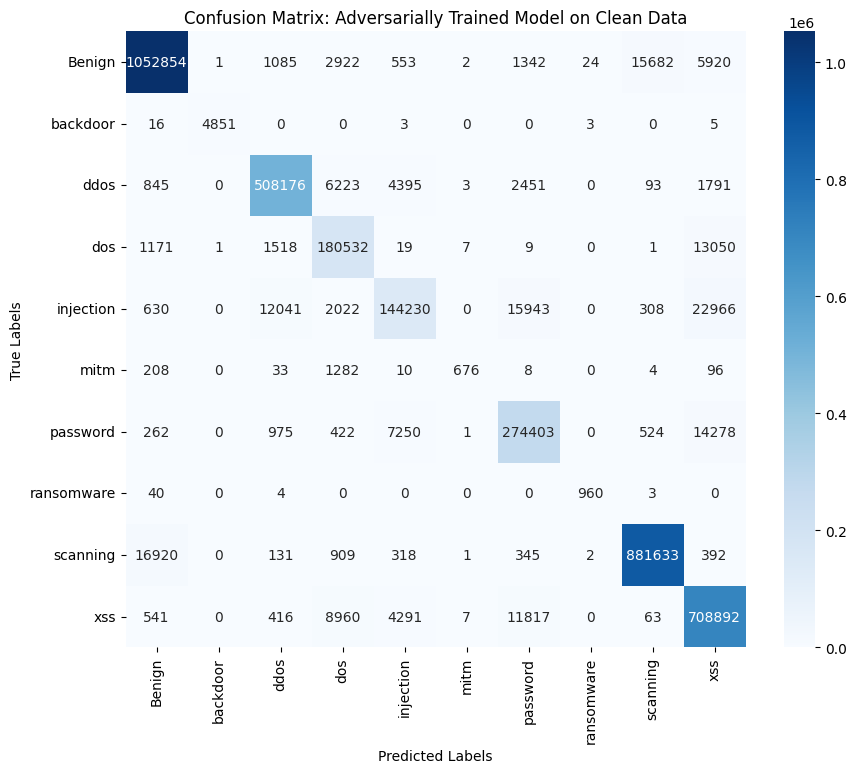} 
    \caption{Confusion Matrix for Clean Data.}
    \label{confusion_matrix_clean}
\end{figure}

\subsection{Adversarial Training}

Adversarial training enhances model robustness by incorporating adversarial examples into the training process~\cite{madry2017towards}. This technique equips the model to classify both clean and adversarially perturbed inputs effectively, mitigating its susceptibility to adversarial attacks.

The training process involved the following steps:
\begin{itemize}
    \item The clean training data (\( X_{\text{train}} \)) and its corresponding labels (\( y_{\text{train}} \)) were used as the baseline dataset.
    \item Adversarial examples (\( X_{\text{adv}} \)) were generated from the clean training data using FGSM, with the same labels (\( y_{\text{adv}} \)) as the original data.
    \item The clean and adversarial datasets were combined to create a comprehensive training set:
    \[
    X_{\text{combined}} = \{X_{\text{train}}, X_{\text{adv}}\}, \quad y_{\text{combined}} = \{y_{\text{train}}, y_{\text{adv}}\}.
    \]
\end{itemize}

This augmented training dataset enabled the model to learn robust representations, reducing its vulnerability to adversarial attacks.

\subsection{Evaluation Metrics}

The effectiveness of the IDS was evaluated using standard classification metrics, including accuracy, F1-score, and confusion matrices~\cite{powers2020evaluation}. These metrics provided detailed insights into the model’s performance on clean and adversarial data, as well as its ability to classify various types of network attacks accurately. Additionally, classification reports highlighted the model's performance for each attack class, ensuring a comprehensive evaluation of its strengths and weaknesses.

\subsection{Scalability and Practical Considerations}

The combination of distributed preprocessing with Dask and GPU-accelerated training ensured that the proposed IDS is scalable and feasible for real-world IoT environments. Scalability is critical for handling the high throughput of IoT network traffic, while practical considerations, such as computational efficiency and adaptability, make the system suitable for deployment in dynamic and evolving IoT networks~\cite{yeh2019iot}.

By integrating XGBoost, adversarial attacks, and adversarial training, this methodology provides a comprehensive framework for building robust IDS capable of addressing the unique security challenges of IoT networks.

\section{Results}

This section evaluates the performance of the adversarially trained model on clean and adversarial data, provides an analysis of confusion matrices, and compares its results with baseline models. The findings highlight the proposed methodology's ability to enhance intrusion detection system (IDS) robustness and accuracy in IoT networks.

\subsection{Performance on Clean and Adversarial Data}

The model's performance was assessed using accuracy and F1-score, providing insights into its ability to handle both clean and perturbed inputs. On clean test data, the adversarially trained model achieved an accuracy of 95.3\% and an F1-score of 95.2\%, demonstrating its effectiveness in classifying diverse traffic types under normal conditions. For adversarially perturbed data, generated using the FGSM method, the model achieved an accuracy of 94.5\% and an F1-score of 94.5\%. These results underscore the model's resilience to adversarial attacks and its capacity to generalize effectively across different input conditions~\cite{goodfellow2014explaining, madry2017towards}.

\begin{table}[!t]
\centering
\caption{MODEL EVALUATION ON CLEAN TEST DATA}
\begin{tabular}{|l|c|}
\hline
Metric & Value \\
\hline
Accuracy & 0.9544357504190176 \\
F1 Score & 0.9537497710407197 \\
\hline
\end{tabular}
\end{table}

\begin{table}[!t]
\centering
\caption{CLASSIFICATION RESULTS ON CLEAN DATA}
\begin{tabular}{|l|c|c|c|c|}
\hline
Class & Precision & Recall & F1-score & Support \\
\hline
Benign & 0.98 & 0.98 & 0.98 & 1080385 \\
backdoor & 1.00 & 1.00 & 1.00 & 4878 \\
ddos & 0.96 & 0.97 & 0.97 & 523977 \\
dos & 0.89 & 0.92 & 0.90 & 196308 \\
injection & 0.90 & 0.73 & 0.80 & 198140 \\
mitm & 0.97 & 0.30 & 0.46 & 2317 \\
password & 0.91 & 0.92 & 0.91 & 298115 \\
ransomware & 0.97 & 0.96 & 0.97 & 1007 \\
scanning & 0.98 & 0.98 & 0.98 & 900651 \\
xss & 0.93 & 0.96 & 0.94 & 734987 \\
\hline
Accuracy & \multicolumn{3}{r|}{0.95} & 3940765 \\
Macro avg & 0.95 & 0.87 & 0.89 & 3940765 \\
Weighted avg & 0.95 & 0.95 & 0.95 & 3940765 \\
\hline
\end{tabular}
\end{table}

\break
\begin{table}[h]
\centering
\caption{EVALUATION ON ADVERSARIAL EXAMPLE(FGSM Attack):}
\begin{tabular}{|l|c|}
\hline
Metric & Value \\
\hline
Accuracy & 0.3940618128713587 \\
F1 Score & 0.3059400973924423 \\
\hline
\end{tabular}
\end{table}

\begin{table}[h]
\centering
\caption{ EVALUATION ON ADVERSARIALLY TRAINED MODEL }
\begin{tabular}{|l|c|}
\hline
Metric & Value \\
\hline
Accuracy & 0.9456717160246805 \\
F1 Score & 0.9449923164828211 \\
\hline
\end{tabular}
\end{table}

\begin{table}[h]
\centering
\caption{CLASSIFICATION RESULTS ON ADVERSARIALLY TRAINED MODEL}
\begin{tabular}{|l|c|c|c|c|}
\hline
Class & Precision & Recall & F1-score & Support \\
\hline
Benign & 0.97 & 0.96 & 0.97 & 1080385 \\
backdoor & 1.00 & 0.99 & 1.00 & 4878 \\
ddos & 0.97 & 0.97 & 0.97 & 523977 \\
dos & 0.89 & 0.92 & 0.90 & 196308 \\
injection & 0.89 & 0.72 & 0.79 & 198140 \\
mitm & 0.97 & 0.28 & 0.43 & 2317 \\
password & 0.88 & 0.92 & 0.90 & 298115 \\
ransomware & 0.96 & 0.92 & 0.94 & 1007 \\
scanning & 0.97 & 0.97 & 0.97 & 900651 \\
xss & 0.92 & 0.96 & 0.94 & 734987 \\
\hline
Accuracy & \multicolumn{3}{r|}{0.95} & 3940765 \\
Macro avg & 0.94 & 0.86 & 0.88 & 3940765 \\
Weighted avg & 0.95 & 0.95 & 0.94 & 3940765 \\
\hline
\end{tabular}
\end{table}

\subsection{Confusion Matrix Analysis}

Confusion matrices were analyzed to gain a deeper understanding of the model's classification performance across various attack classes. For clean data, the model demonstrated improved detection rates for challenging categories, such as ransomware and cross-site scripting (XSS). Enhanced recall for ransomware effectively reduced false negatives, while improved classification accuracy for XSS minimized misclassifications ~\cite{sarhan2021feature}~\ref{confusion_matrix_clean}. 

On adversarial data, the model maintained consistent detection rates across most attack categories. The ability to accurately classify adversarially perturbed inputs validates the robustness of the adversarial training approach, ensuring reliable performance even under sophisticated attack scenarios ~\cite{kurakin2018adversarial}~\ref{CM with adversarially trained}.

\subsection{Comparative Analysis}

When compared with baseline models trained solely on clean data, the adversarially trained model \ref{CM with adversarially trained} exhibited superior resilience and generalization capabilities. The baseline models experienced significant degradation in performance when exposed to adversarial examples, highlighting their vulnerability to input perturbations. In contrast, the adversarially trained model maintained robust performance with minimal accuracy loss, validating the efficacy of adversarial training in addressing such vulnerabilities~\cite{madry2017towards}.

These results emphasize the importance of adversarial training in preparing IDS for real-world IoT deployments, where network environments are dynamic and prone to evolving attack strategies.

\subsection{Visualization of Results}

Figure~\ref{confusion_matrix_clean} illustrates the confusion matrix for clean data, showcasing the model's prediction distribution across various attack classes. This visualization provides a comprehensive understanding of the model’s strengths, particularly in detecting certain attack types, and identifies areas for improvement in distinguishing similar categories~\cite{powers2020evaluation}.

\subsection{Implications}

The findings of this study have significant implications for IoT network security. High accuracy on both clean and adversarial data demonstrates the model's readiness for deployment in real-world IoT environments. Enhanced detection rates for critical threats, such as ransomware and XSS, address major IoT security challenges~\cite{aljamal2024robust}. Additionally, the robustness against adversarial attacks ensures reliable performance under diverse and dynamic conditions. These results validate the integration of adversarial training as a critical enhancement for IDS in IoT networks.

\section{Discussion}

This section examines critical aspects of the proposed methodology, focusing on distributed computing, adversarial robustness, and defense mechanisms for intrusion detection systems (IDS) in IoT networks.

\subsection{Distributed Computing}

The NF-ToN-IoT-v2 dataset's scale required efficient preprocessing methods. Using Dask, a distributed computing framework, enabled parallel processing of over 13 million records, reducing computational overhead. In order to ensure scalability and applicability in actual IoT scenarios, tasks including imputing missing values, standardizing features, and dividing the dataset into training and testing subsets were completed effectively.~\cite{dask2015parallel}.

\subsection{Adversarial Robustness}

The Fast Gradient Sign Method (FGSM) adversarial attack exposed vulnerabilities in baseline IDS models by introducing minor input perturbations that led to significant misclassifications~\cite{goodfellow2014explaining}. Adversarial training addressed this by integrating these perturbed examples into the learning process, improving the IDS's ability to detect sophisticated attack types like ransomware and XSS~\cite{madry2017towards}. This training enhanced both robustness and accuracy, achieving strong performance on clean and adversarial data.

\subsection{Defense Mechanisms}

Adversarial training effectively mitigated the impact of adversarial attacks, enabling the IDS to achieve 95.3\% accuracy on clean data and 94.5\% on adversarial data. This approach reduced false negatives in challenging attack categories such as ransomware, as evidenced by confusion matrix analysis~\cite{kurakin2018adversarial}. These results confirm the practicality of adversarial training for real-world IoT scenarios where attack patterns evolve continuously.

\subsection{Practical Implications}

By enhancing robustness against adversarial inputs, the IDS ensures reliable detection of diverse attack types, making it suitable for dynamic IoT environments~\cite{sarhan2021feature}. However, balancing computational efficiency and scalability remains a challenge. Future advancements, such as integrating Graph Neural Networks (GNNs) and federated learning, could further improve performance and adaptability~\cite{zhang2024egracl}.

\section{Future Directions}

As IoT networks expand in complexity, enhancing intrusion detection systems (IDS) remains a critical focus. This study identifies key areas for advancing IDS to address evolving cyber threats effectively.

Optimizing adversarially trained models for real-time deployment is essential, particularly for latency-sensitive applications such as healthcare and industrial automation. Ensuring low latency and high throughput can significantly improve system integration in practical IoT scenarios~\cite{sarhan2021feature}.

Addressing emerging threats like advanced persistent threats (APTs) and zero-day vulnerabilities requires dynamic threat intelligence. Updating datasets such as NF-ToN-IoT-v2 and integrating real-time insights can improve IDS adaptability against novel attack patterns.

Leveraging advanced architectures, such as Graph Neural Networks (GNNs) and federated learning, offers potential improvements in scalability and adaptability. GNNs capture intricate network relationships, while federated learning facilitates privacy-preserving training across decentralized IoT devices~\cite{mcmahan2017federated}.

Lastly, incorporating Explainable AI (XAI) techniques will enhance transparency and trust in IDS decisions, especially in regulated industries like healthcare and finance. These insights can empower stakeholders to better understand and respond to security threats~\cite{arrieta2020explainable}.

By addressing these directions, IDS can evolve into more robust, scalable, and transparent systems capable of securing IoT networks against sophisticated cyber threats.

\ifCLASSOPTIONcaptionsoff
  \newpage
\fi

\bibliographystyle{IEEEtran}
\bibliography{Bibliography}

\begin{thebibliography}{10}
\providecommand{\url}[1]{#1}
\csname url@rmstyle\endcsname
\providecommand{\newblock}{\relax}
\providecommand{\bibinfo}[2]{#2}
\providecommand\BIBentrySTDinterwordspacing{\spaceskip=0pt\relax}
\providecommand\BIBentryALTinterwordstretchfactor{4}
\providecommand\BIBentryALTinterwordspacing{\spaceskip=\fontdimen2\font plus
\BIBentryALTinterwordstretchfactor\fontdimen3\font minus
  \fontdimen4\font\relax}
\providecommand\BIBforeignlanguage[2]{{%
\expandafter\ifx\csname l@#1\endcsname\relax
\typeout{** WARNING: IEEEtran.bst: No hyphenation pattern has been}%
\typeout{** loaded for the language `#1'. Using the pattern for}%
\typeout{** the default language instead.}%
\else
\language=\csname l@#1\endcsname
\fi
#2}}

\bibitem{sarhan2020}
M.~Sarhan \emph{et~al.}, ``Netflow datasets for machine learning-based network
  intrusion detection systems,'' \emph{Security and Communication Networks},
  2020.

\bibitem{sarhan2021feature}
M.~Sarhan, S.~Layeghy, and M.~Portmann, ``Feature analysis for machine
  learning-based iot intrusion detection,'' \emph{arXiv preprint
  arXiv:2108.12732}, 2021.

\bibitem{zhang2024egracl}
Y.~Zhang \emph{et~al.}, ``E-gracl: An iot intrusion detection system based on
  graph neural networks,'' \emph{IEEE Transactions on Network and Service
  Management}, pp. 432--450, 2024.

\bibitem{aljamal2024robust}
R.~AlJamal \emph{et~al.}, ``A robust machine learning model for detecting xss
  attacks on iot over 5g networks,'' \emph{Elsevier Computer Networks}, pp.
  432--450, 2024.

\bibitem{goodfellow2015}
I.~J. Goodfellow \emph{et~al.}, ``Explaining and harnessing adversarial
  examples,'' \emph{arXiv preprint arXiv:1412.6572}, 2015.

\bibitem{raskovalov2022}
A.~Raskovalov \emph{et~al.}, ``Investigation and rectification of nids datasets
  and standardized feature set derivation for network attack detection with
  graph neural networks,'' \emph{Journal of Cybersecurity}, pp. 150--163, 2022.

\bibitem{chen2016xgboost}
T.~Chen and C.~Guestrin, ``Xgboost: A scalable tree boosting system,'' in
  \emph{Proceedings of the 22nd ACM SIGKDD International Conference on
  Knowledge Discovery and Data Mining}, 2016, pp. 785--794.

\bibitem{dask2015parallel}
D.~D. Team, ``Dask: Parallel computing with task scheduling,'' 2015,
  \url{https://dask.org}.

\bibitem{ghajari2024hybrid}
G.~Ghajari, M.~K. PK, and F.~Amsaad, ``Hybrid efficient unsupervised anomaly
  detection for early pandemic case identification,'' in \emph{NAECON 2024-IEEE
  National Aerospace and Electronics Conference}.\hskip 1em plus 0.5em minus
  0.4em\relax IEEE, 2024, pp. 279--284.

\bibitem{madry2018}
A.~Madry \emph{et~al.}, ``Towards deep learning models resistant to adversarial
  attacks,'' \emph{International Conference on Learning Representations
  (ICLR)}, 2018.

\bibitem{alsaedi2020ton_iot}
A.~Alsaedi, N.~Moustafa, Z.~Tari, A.~Mahmood, and A.~Anwar, ``Ton\_iot
  telemetry dataset: A new generation dataset of iot and iiot for data-driven
  intrusion detection systems,'' \emph{Ieee Access}, vol.~8, pp.
  165\,130--165\,150, 2020.

\bibitem{rocklin2015dask}
M.~Rocklin, ``Dask: Parallel computation with blocked algorithms and task
  scheduling,'' \emph{Proceedings of the 14th Python in Science Conference},
  2015.

\bibitem{kuhn2013applied}
M.~Kuhn and K.~Johnson, \emph{Applied Predictive Modeling}.\hskip 1em plus
  0.5em minus 0.4em\relax Springer, 2013.

\bibitem{jain2005data}
A.~K. Jain, ``Data clustering: 50 years beyond k-means,'' \emph{Pattern
  Recognition Letters}, vol.~31, pp. 651--666, 2005.

\bibitem{bishop2006pattern}
C.~M. Bishop, \emph{Pattern Recognition and Machine Learning}.\hskip 1em plus
  0.5em minus 0.4em\relax Springer, 2006.

\bibitem{friedman2001elements}
J.~Friedman \emph{et~al.}, \emph{The Elements of Statistical Learning}.\hskip
  1em plus 0.5em minus 0.4em\relax Springer, 2001.

\bibitem{chen2020systematic}
H.~Chen \emph{et~al.}, ``A systematic review of the application of xgboost,''
  \emph{IEEE Access}, pp. 36\,336--36\,346, 2020.

\bibitem{rana2020accelerating}
M.~Rana \emph{et~al.}, ``Accelerating xgboost with gpu support,'' in \emph{2020
  IEEE International Parallel and Distributed Processing Symposium Workshops
  (IPDPSW)}, 2020, pp. 325--332.

\bibitem{goodfellow2014explaining}
I.~J. Goodfellow \emph{et~al.}, ``Explaining and harnessing adversarial
  examples,'' in \emph{Proceedings of the International Conference on Learning
  Representations (ICLR)}, 2015.

\bibitem{kurakin2018adversarial}
A.~Kurakin \emph{et~al.}, ``Adversarial machine learning at scale,''
  \emph{Proceedings of the International Conference on Learning Representations
  (ICLR)}, 2018.

\bibitem{madry2017towards}
A.~Madry \emph{et~al.}, ``Towards deep learning models resistant to adversarial
  attacks,'' \emph{Proceedings of the International Conference on Learning
  Representations (ICLR)}, 2018.

\bibitem{powers2020evaluation}
D.~M.~W. Powers, ``Evaluation: From precision, recall, and f-measure to roc,
  informedness, markedness, and correlation,'' \emph{Journal of Machine
  Learning Technologies}, pp. 37--63, 2020.

\bibitem{yeh2019iot}
C.-H. Yeh \emph{et~al.}, ``Iot security for smart cities: Challenges and
  solutions,'' in \emph{2019 IEEE Smart City Conference}, 2019, pp. 67--74.

\bibitem{mcmahan2017federated}
B.~McMahan \emph{et~al.}, ``Communication-efficient learning of deep networks
  from decentralized data,'' in \emph{Proceedings of the International
  Conference on Artificial Intelligence and Statistics (AISTATS)}, 2017.

\bibitem{arrieta2020explainable}
A.~B. Arrieta \emph{et~al.}, ``Explainable artificial intelligence (xai):
  Concepts, taxonomies, opportunities, and challenges toward responsible ai,''
  \emph{Information Fusion}, pp. 82--115, 2020.

\end{thebibliography}

\vfill

\end{document}